\documentclass[aps, twocolumn, superscriptaddress]{revtex4} 

\usepackage[switch]{lineno}
\usepackage{empheq}
\usepackage[parfill]{parskip}
\usepackage[active]{srcltx}
\usepackage{color}
\usepackage{array}
\usepackage{booktabs}
\usepackage{amsfonts}
\usepackage{graphicx}
\usepackage{dcolumn}
\usepackage{bm}
\usepackage{epstopdf}
\usepackage{xcolor}
\usepackage{textcmds}
\usepackage{textcomp}
\usepackage{soul}
\usepackage{amsbsy}
\usepackage{mathrsfs} 
\usepackage{amsmath}
\usepackage{bigints} 
\usepackage{amsthm,amssymb} 
\usepackage[utf8]{inputenc}
\usepackage[english]{babel}
\usepackage[shortlabels]{enumitem}
\usepackage{float}
\usepackage{mathrsfs,bigints,mathtools,dsfont}
\usepackage[colorlinks=true,linkcolor=blue,citecolor=blue]{hyperref}%
\usepackage[toc]{appendix}

\begin{document}

\setlength{\parindent}{0.5cm}

\title{A solvable two-dimensional swarmalator model}

\author{Kevin O'Keeffe}
\email{Corresponding to: kevin.p.okeeffe@gmail.com}
\noaffiliation

\author{Gourab Kumar Sar}
\email{mr.gksar@gmail.com}
\affiliation{Physics and Applied Mathematics Unit, Indian Statistical Institute, 203 B. T. Road, Kolkata 700108, India}

\author{Md Sayeed Anwar}
\affiliation{Physics and Applied Mathematics Unit, Indian Statistical Institute, 203 B. T. Road, Kolkata 700108, India}

\author{Joao U. F. Lizárraga}
\affiliation{Instituto de Física Gleb Wataghin, Universidade Estadual de Campinas, Unicamp 13083-970, Campinas, São Paulo, Brazil}

\author{Marcus A. M. de Aguiar}
\affiliation{Instituto de Física Gleb Wataghin, Universidade Estadual de Campinas, Unicamp 13083-970, Campinas, São Paulo, Brazil}

\author{Dibakar Ghosh}
\affiliation{Physics and Applied Mathematics Unit, Indian Statistical Institute, 203 B. T. Road, Kolkata 700108, India}

\begin{abstract}
\hspace{1 cm}  (Received XX MONTH XX; accepted XX MONTH XX; published XX MONTH XX) \\ 

Swarmalators are oscillators that swarm through space as they synchronize in time. Introduced a few years ago to model many systems which mix synchrony with self-assembly, they remain poorly understood theoretically. Here we obtain the first analytic results on swarmalators moving in two-dimensional (2D) plane by enforcing periodic boundary conditions; this simpler topology allows expressions for order parameters, stabilities, and bifurcations to be derived exactly. We suggest some future directions for swarmalator research and point out some connections to the Kuramoto model and the Vicsek model from active matter; these are intended as a call-to-arms for the sync community and other researchers looking for new problems and puzzles to work on.
\\

\noindent
DOI: XXXXXXX
\end{abstract}

\maketitle

\section{Introduction}
Swarmalators are mobile oscillators whose internal phases couple to their external motions \cite{o2017oscillators}. They are toy models for the many systems in which sync and swarming interact, such as biological microswimmers \cite{riedel2005self,quillen2021metachronal}, magnetic domain walls \cite{hrabec2018velocity}, migratory cells \cite{riedl2022synchronization}, forced colloids \cite{yan2012linking,zhang2023spontaneous,liu2021activity,li2018spatiotemporal} and robotic swarms \cite{ahern2022unifying,adams2023king,barcis2019robots,barcis2020sandsbots,monaco2020cognitive}.

\begin{figure}
    \centering
    \includegraphics[width=\columnwidth]{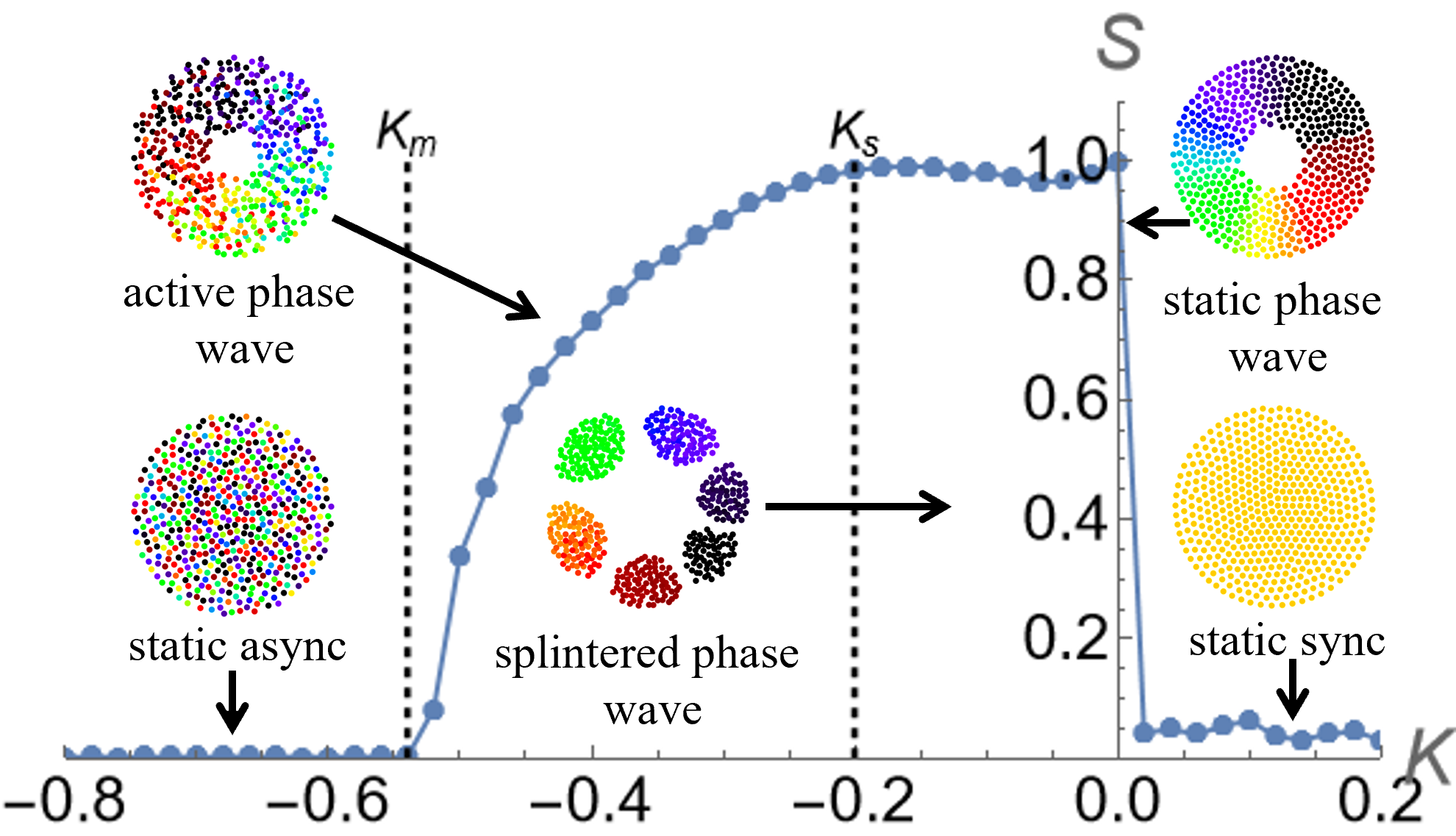}
    \caption{Rainbow order parameter $S(K)$ of the 2D swarmalator model Eqs.~\eqref{uv1},~\eqref{uv2}. Insets show the collective states of the model where swarmalators are represented as colored dots in the $(x,y)$ plane where the color denotes the swarmalators phase. Static async: $(J,K) = (0.5,-0.8)$, active phase wave $(J,K) = (0.5,-0.4)$, splintered phase wave $(J,K) = (0.5,-0.1)$, static phase wave $(J,K) = (0.5,0)$, static sync $(J,K) = (0.5,0.2)$. In the three static states, swarmalators do not move in space or phase. In the splintered phase wave, each colored chunk is a vortex: the swarmalators oscillate in both space and phase. In the active phase wave, the oscillations are excited into rotations. The swarmalators split into two groups with counter-rotate in $x$ and $\theta$.} 
    \label{order-parameters-2D}
\end{figure}

This paper is part of a research program \cite{o2022collective,yoon2022sync,o2022swarmalators} aimed to develop a theory for swarmalators. Our north star goal is to find the `right' model for swarmalators -- like the Kuramoto model from sync studies or the Ising model from statistical physics -- which captures universal behavior but is also solvable. Our hope is the swarmalator model in \cite{o2017oscillators} could fill this role, as it imitates real-world swarmalators and appears tractable. Figure~\ref{order-parameters-2D} shows its collective states and bifurcation structure by plotting the rainbow order parameter $S :=  \max (S_+, S_-)$ where $S_{\pm} = |W_{\pm}| = |\langle e^{i (\phi_i \pm \theta_i )} \rangle |$ and $\phi_i = \arctan(y_i/x_i)$ is the spatial angle (states with $S_+ > S_-$ and $S_- < S_+$ are equally likely; we define $S$ as the max of these to break the degeneracy). At $K = K_m$, $S$ spikes up from zero as static async melts into an active phase wave where swarmalators run in a space-phase vortex. At a later $K = K_s$, $S$ starts to decline as the the vortex splits into a ring of mini-vortices called the splintered phase wave. Our aim is to predict the melting point $K_m$, the splitting point $K_s$, and ideally the stable branches of $S(K)$. We see these as first steps towards building a theory.

Take the melting point $K_m$. How could you find it? You might think you could use linearization to find the stability spectrum of the asynchronous state $\rho(\mathbf{x},\theta) = (4 \pi^2)^{-1} H(r-\mathcal{R})$, where $H(r)$ is Heaviside's function and $\mathcal{R}$ is the disk's radius. But the sharp edges of the disk come with a hard mathematical wall. They lead to non-standard eigenvalue equations of form $\lambda b (\mathbf{x}) = H(r-\mathcal{R}) \int b(\mathbf{y}) / |\mathbf{x}-\mathbf{y}| d \mathbf{y} + \delta(r-\mathcal{R}) \int b(\mathbf{y})  (\mathbf{y}-\mathbf{x}) d \mathbf{y} $ for which standard tools falter. Convolutions like $\int b(\mathbf{y})/|\mathbf{x}-\mathbf{y}| d \mathbf{y}$ are already hard to deal with. The Heaviside and delta functions out front make things even worse.

Another idea might be to find $K_m$ using Kuramoto's classic self-consistency method \cite{kuramoto2003chemical,strogatz2000kuramoto}. But that path has its own blocker. The reason is technical, but the essence is that in 1D, a zero divergence condition is easy, but in 3D, it is hard. We'll sketch why this is the sticking point below, but don't fixate on getting every detail right. Just try to absorb the big picture.

Kuramoto's trick to find the critical coupling $K_c$ in his model is this: write the sync order parameter $R := |\langle e^{i \theta} \rangle| = \int e^{i \theta} \rho(\theta, \omega) g(\omega) d \theta d\omega $ in terms of itself by finding how the density of the sync state $\rho(\theta,\omega)$ depends on $R$ (in the integral $g(\omega)$ is the distribution of natural frequencies). This $\rho(\theta,\omega)$ satisfies the stationary 1D continuity equation $\nabla \cdot (v(R) \rho) = 0$ where the velocity depends on $R$ \footnote{1D because a population of oscillators with $\theta \in \mathbb{S}^1$ may be thought of as a fluid moving in $S^1$}. Here's the crucial part: in 1D, the $\nabla$ operator is just a single derivative $\partial_{\theta}$ which when zero implies a constant: $\nabla \cdot ( v \rho) = 0 \Rightarrow \ \partial_{\theta}(v \rho) \Rightarrow \rho = 0 \Rightarrow \rho = C / v(R)$ for some constant $C$. Thus, we have our prize $\rho$ in terms of $R$ and can sub this back into $R = |\int e^{i \theta} \rho(\theta, \omega) g(\omega) d \theta d\omega|$. Then you take a limit $R \rightarrow 0$ to find the target $K_c$. (By definition, $K_c$ is when $R$ jumps from zero.) 

Can we adopt this procedure for swarmalators? Find their density $\rho(\mathbf{x},\theta)$ in terms of its order parameter $S$ and invoke self-consistency? Not so easily it turns out, because $\rho(\mathbf{x}, \theta)$, being defined over three dimensions $(\mathbf{x}, \theta) \in \mathbb{R}^2 \times \mathbb{S}^1$, is not so easily trapped. The steady state condition $\nabla \cdot (v \rho) = 0$ does not reduce to $\rho v = C$  \footnote{Recall a vector field $\mathbf{F}$ with zero divergence may be written as $\mathbf{F} = \nabla \times \mathbf{A}$, but that doesn't help us here}. It becomes an nonlinear integro-differential equation of form: $\rho . \nabla v + v_x \nabla_x \rho + v_{\theta} \partial_{\theta} \rho = 0$ where the velocities are convolutions over the density $v_x ,v_{\theta} = \int F(\cdot) \rho(\cdot)$ for some kernels $F$. On top of that, the analogue of the $R = |\int e^{i \theta} \rho(\theta,\omega,t)|$ integral for the swarmalator order parameter $S$ becomes a monster: a 3D contour integral over the support of the active phase wave $S = \max_{\pm} |\int_{\Gamma(r, \phi, \theta)} e^{i ( \phi \pm \theta) } \rho(S_+, S_-, r,\phi,\theta) dr d \phi d \theta |$. Note we have switched to polar coordinates here, and defined $\Gamma$ as the support of the active phase. Note also that the active phase wave is the relevant density here because that's the state into which the async state bifurcates (Fig.~\ref{order-parameters-2D}).

To repeat the big picture: to use Kuramoto's self-consistency approach for swarmalators, you need the steady state $\rho(\mathbf{x},\theta)$ in terms of the order parameter $S$, but in 3D that requires solving a nonlinear integro-differential equation.

Where does that leave us? We were brainstorming ways to pin down the melting point $K_m$, which is a stepping stone towards building a swarmalator theory. But our search hit two walls: The 3D flow, which complicates a Kuramoto-style, self-consistency approach, and the compact support of the async state, which makes for a thorny stability analysis. 

This paper tries to make analytic headway on these problems by confining the swarmalators' movements to a 2D plane with periodic boundary conditions (as opposed to the open plane $x \in \mathbb{R}^2$ of the original 2D swarmalator model). This leads to simpler analogues of four of the five swarmalator states in Fig.~\ref{states} (the splintered phase wave is absent) which allow us to calculate the melting point $K_m$ and the rainbow order parameter $S(K)$ for both identical and non-identical swarmalators. 

Finally, we close with an argument for why these questions about swarmalators are worth asking for the sync community, and also for science more generally.
\begin{figure*}
    \centering
    \includegraphics[width= 2\columnwidth]{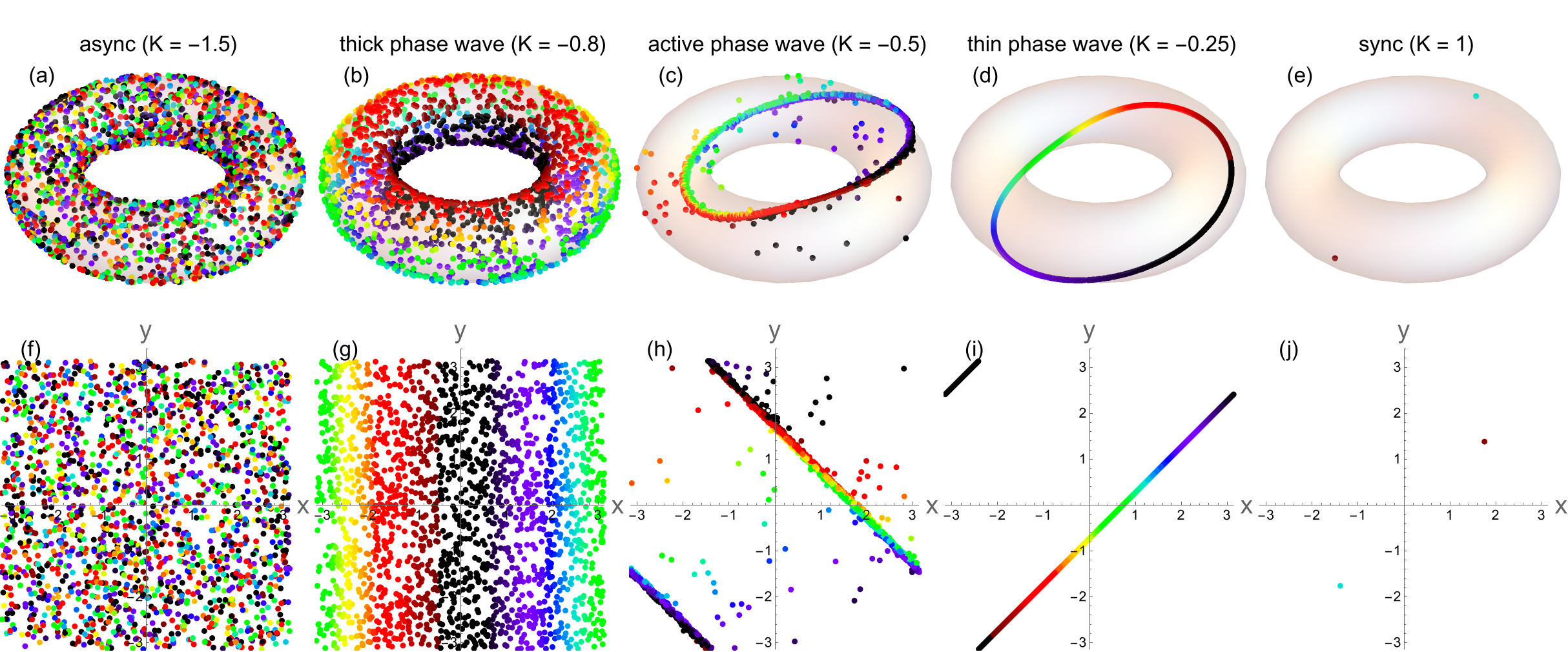}
    \caption{Collective states of the model Eqs.~\eqref{x1}-\eqref{t1}. Top row, swarmalators are depicted as dots on the torus $(x_i, y_i)$ with the color indicating the phase $\theta_i$. Bottom row, same information in the $(x, y)$ plane. $J=1$ here. (a, f) async for $K=-1.5$, (b, g) thick phase wave for $K=-0.8$, (c, h) active phase wave for $K=-0.5$, (d, i) thin phase wave for $K=-0.25$, (e, j) sync for $K=1$. Simulation details: step-size $dt=0.5$, final time $T=200$, and total number of swarmalators $N=2000$ using an RK4 solver.} 
    \label{states-torus}
\end{figure*}
\section{Model}
The swarmalator model is 
\begin{align}
&\dot{\mathbf{x}}_i =  \frac{1}{N} \sum_{j=1}^N \Big[ \mathbf{I}_{\mathrm{att}}(\mathbf{x}_j - \mathbf{x}_i)F(\theta_j - \theta_i)  - \mathbf{I}_{\mathrm{rep}} (\mathbf{x}_j - \mathbf{x}_i) \Big],   \\ 
& \dot{\theta_i} = \omega_i +  \frac{K}{N} \sum_{j=1}^N H_{\mathrm{\mathrm{sync}}}(\theta_j - \theta_i) G(\mathbf{x}_j - \mathbf{x}_i).
\end{align}
The spatial dynamics encode self-assembly which depends on sync, the phase dynamics sync which depends on self-assembly. Looking at $\dot{\mathbf{x}}_i$, we see a combination of pairwise repulsion $I_{rep}(\mathbf{x}_j - \mathbf{x}_i)$ and attraction $I_{att}(\mathbf{x}_j - \mathbf{x}_i)$ modified by phase differences $F(\theta_j -\theta_i)$. In $\dot{\theta}_i$, we see a natural frequency $\omega_i$ and a sync term $H_{sync}(\theta_j - \theta_i)$ modified by a distance kernel $G(\mathbf{x}_j - \mathbf{x}_i)$.  

The instance of the model studied in \cite{o2017oscillators} was:
\begin{align}
&\dot{\mathbf{x}}_i = \frac{1}{N} \sum_{ j \neq i}^N \Bigg[ \frac{\mathbf{x}_j - \mathbf{x}_i}{|\mathbf{x}_j - \mathbf{x}_i|} \Big(1+J \cos(\theta_j - \theta_i)  \Big) -   \frac{\mathbf{x}_j - \mathbf{x}_i}{ | \mathbf{x}_j - \mathbf{x}_i|^2}\Bigg] , \label{uv1} \\ 
& \dot{\theta_i} = \frac{K}{N} \sum_{j \neq i}^N \frac{\sin(\theta_j - \theta_i )}{|\mathbf{x}_j - \mathbf{x}_i|}  \label{uv2}.
\end{align}
\noindent
Notice, this is for identical swarmalators $\omega_i = \omega$ with $\omega=0$ without loss of generality by a change of frame.

What makes this model difficult to solve is the power law kernels -- in particular, the hard shell $1/|\mathbf{x}|^p$ terms. The cosine and sine terms on the other hand are relatively easy to deal with. In the $N \rightarrow \infty $ limit, they may be expressed in terms of a finite number of Fourier harmonics of the density. To see this, consider the Kuramoto model with identical frequencies $\dot{\theta}_i = \omega + K/N \sum_j \sin(\theta_j - \theta_i)$ with density $f(\theta)$. In the continuum limit, sums of sines of pairwise phases simplify like $\sum_j \sin(\theta_j - \theta_i) \rightarrow \int \sin(\theta' - \theta) f(\theta') d \theta' = R_1 \sin(\phi_1 -\theta)$ where $R_1 e^{i \phi_1} := \int e^{i \theta'} f(\theta') d \theta'$ is the first harmonic. The benefit of this is when $v(\theta) = \omega + K R \sin(\phi -\theta)$ is plugged into the continuity equation $\dot{f} + \partial_{\theta}(v f) =0$, and $f$ is expanded in a Fourier series $f(\theta) = (2 \pi)^{-1}(1 +\sum_n R_n e^{i \phi_n} e^{i n \theta} + c.c.)$, the amplitude equations $\dot{R}_n$ are relatively simple; $f$, which has an infinite number of $R_n$, interacts with $v$, which has just one $R_1$.

The point is if you can express the velocity fields $v$ in term of Fourier harmonics of the density, calculations tend to get easier. This is the feature we want to bake into our swarmalator model. So we replace the power law kernels  with simple trigonometric terms by restricting the swarmalators movements to a plane with periodic boundary conditions which we think of as the torus $(x_i, y_i) \in (\mathbb{S}^1, \mathbb{S}^1)$ under a suitable length scale:
\begin{align}
    \dot{x}_i &= \frac{J}{N} \sum_j \sin(x_j - x_i) \cos(\theta_j - \theta_i) \label{x1}, \\
    \dot{y}_i &= \frac{J}{N} \sum_j \sin(y_j - y_i)\cos(\theta_j - \theta_i) \label{y1}, \\
    \dot{\theta}_i &= \frac{K}{N} \sum_j \sin(\theta_j - \theta_i) \Big(\cos(x_j - x_i) + \cos(y_j - y_i) \Big). \label{t1} 
\end{align}
Notice we set the spatial attraction to $I_{att}(x) = \sin(x)$ and turned off the hard shell repulsion $I_{rep}(x) = 0$ for simplicity. Originally we replaced the distance kernel $G(\mathbf{x}) =1/|\mathbf{x}|$ with $G(x,y) = 2 + \hat{d}(x,y)$ where $\hat{d}$ is a new distance metric on the torus: $ \hat{d}(x,y) = \cos x + \cos y$, which enforces $G$ to decrease monotonically on $[2,0]$ (as we want, since it represents spatial decay). But we found qualitatively similar phenomena with the simpler choice $G(x,y) = \cos x + \cos y$, so we used that instead. The same was true of the $H(\theta) = 1 + J \cos\theta$ term, which we replaced with $J \cos \theta$.

This much simpler swarmalator model captures behavior of the 2D model yet may be solved for both identical and non-identical swarmalators (where there are natural frequencies $(u,v,w) = (u_i, v_i, w_i)$ added in before the sums in Eqs.~\eqref{x1}-\eqref{t1}). It is a clean generalization of the 1D swarmalator model \cite{o2022collective,yoon2022sync} $\dot{x}_i = J/N \sum \sin(x_j-x_j) \cos(\theta_j-\theta_i), \dot{\theta}_i = K/N \sum \sin(\theta_j-\theta_j) \cos(x_j- x_i)$ and so can be analyzed with the same techniques.

\section{Identical swarmalators}
Figure~\ref{states-torus} shows this model produces analogues of the collective states of the original model (excepting the splintered phase wave). 

\textit{Static async}. Swarmalators are uniformly in space and phase and so are fully incoherent.

\textit{Thick phase wave}. Swarmalators are distributed uniformly over $(x,y)$ with their phases perfectly correlated with either $x$ or $y$. $\theta_i = \pm x_i + C, \theta_i = \pm y_i + C$ where $C$ is any constant. Numerics indicated each member of this $\pm$ family is equally likely.

\textit{Thin phase wave}. Now all three coordinates are perfectly correlated $x_i = \pm y_i = \pm \theta_i + C$ (with each $\pm$ variation equally likely). 

\textit{Active phase wave}. Like the thin phase wave but now the swarmalator execute shear flow around the center correlation line $x_i = \pm y_i = \pm \theta_i$. This state is in fact a long transient and eventually decays to the thin phase wave, but we tabulate it here as a separate state so that future researchers won't be confused by it (as we were; it took some work to diagnose it as a transient) 

\textit{Sync}. Swarmalators collapse to fully synchronous point masses separated $\pi$ unit apart $x_i, y_i, \theta_i = C + n \pi$. where $n = 0,1$. For some initial conditions, only one cluster forms ($n=0$), for others two clusters form ($n=1$). The one/two cluster structure arises from a $\pi$-symmetry \cite{yoon2022sync} in the governing equations: shifting $x_j - x_i, y_j - y_i, \theta_j - \theta_i$ by $\pi$ does not change the equations of motion.

Supplementary Movie 1 shows the evolution of the states. Next we analyze their bifurcations. Mathematica files of our analysis can be found here \footnote{https://github.com/Khev/swarmalators/tree/master/2D/on-torus}.

\textbf{Analysis of sync}. We analyze the stability of the 1-cluster state without loss of generality for which
\begin{align}
    x_i = C_x, \\
    y_i = C_y, \\
    \theta_i = C_{\theta},
\end{align}
are the fixed points. $C_{(\cdot)}$ is a constant determined by the initial conditions. Linearizing the governing equations yields
\begin{equation}
    M = \left[ 
\begin{array}{ccc} 
  X_x & X_y & X_{\theta} \\ 
  Y_x & Y_y & Y_{\theta} \\ 
  T_x & T_y & T_{\theta} \\ 
\end{array} 
\right] 
\end{equation}
where $(X_x)_{ij} = \frac{\partial \dot{x}_i}{\partial x_j}, (Y_x)_{ij} = \frac{\partial \dot{y}_i}{\partial x_j}, TX_x)_{ij} = \frac{\partial \dot{\theta}_i}{\partial x_j}$ and so on. Evaluating $M$ at the fixed points results in 
\begin{equation}
    M = \left[ 
\begin{array}{ccc} 
  A(J) & 0 & 0 \\ 
  0 & A(J) & 0 \\ 
  0 & 0 & A(2K) \\ 
\end{array} 
\right] 
\end{equation}
where
\begin{equation}
A := \begin{bmatrix}
    - \frac{N-1}{N}   & \frac{1}{N}  & \dots & \frac{1}{N}\\
    \frac{1}{N}  & - \frac{N-1}{N} &  \dots & \frac{1}{N} \\
    \hdotsfor{4} \\
    \frac{1}{N}  & \frac{1}{N}  & \dots & - \frac{N-1}{N} 
\end{bmatrix} \label{A}
\end{equation}
Exploiting the block structure of $M$ we find the eigenvalues $\lambda$ are
\begin{align}
    & \lambda_1 = - J, \\
    & \lambda_2 = - 2K, \\
    & \lambda_3 = 0,
\end{align}
with multiplicities $2N-2, N-1, 3$ (which sum to a total of $3N$ as required, since the Jacobean has dimension $(3N)^2$, one $N $ for each of three coordinates $x,y,\theta$). These show the state is stable for $J,K>0$ and destabilize via a saddle node bifurcation. The zero eigenvalues stem from the model's rotational invariance.

\textbf{Analysis of thin phase wave}. We analyze the thin phase wave of form
\begin{align}
    x_i = (i-1)/N  + C_x, \\
    y_i = (i-1)/N  + C_y, \\
    \theta_i = (i-1)/N + C_{\theta},
\end{align}
for $i = 0, \dots, N$ without loss of generality. We again compute the $M$ and exploit its block strucutre.  The calculation mirrors that in \cite{o2022collective} and eventually yields
\begin{align}
    & \lambda_1 = - \frac{J}{2}, \\
    & \lambda_2 = - \frac{J}{4}, \\
    & \lambda_3 = -J-2K, \\
    & \lambda_4 = \frac{1}{8} \left(-J - 2 K - \sqrt{J^2 + 68 J K + 4 K^2}\right) .
\end{align}
This tells us the state is stable for $-J - 2K < 0$
\begin{align}
    K_c = -\frac{J}{2}, \label{eq.21}
\end{align}
which generalizes the $K_c$ of the 1D model \cite{o2022collective}. The bifurcation is a saddle node and a subcritical Hopf bifurcation, meaning an unstable limit cycle is born at $K _c$. 
\begin{figure}[t!]
    \centering
    \includegraphics[width= \columnwidth]{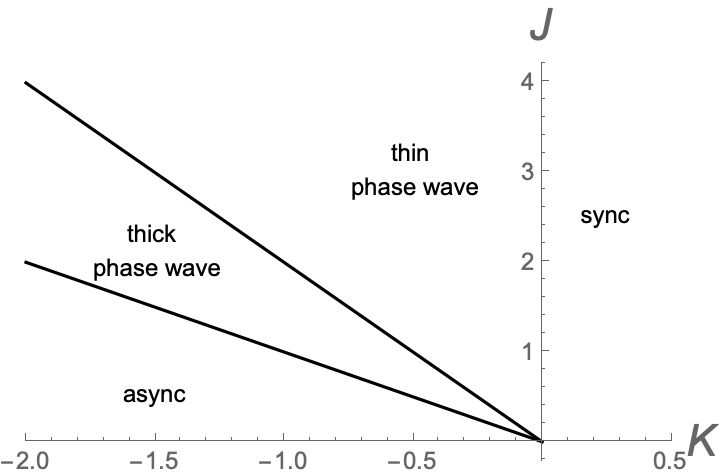}
    \caption{Bifurcation diagram for identical swarmalators governed by model Eqs.~\eqref{x1}-\eqref{t1}. Thick black lines are theoretical predictions. See main text for details.} 
    \label{bif-diagram-model1}
\end{figure}

\textbf{Analysis of async}. This state is best analyzed in the continuum limit where the density of oscillators is $\rho(x,y,\theta) = (2\pi)^{-3}$. We also switch to sum/difference coordinates, 
\begin{align}
    x_{\pm} = x \pm \theta, \\
    y_{\pm} = y \pm \theta.
\end{align}
The density obeys the continuity equation
\begin{align}
\dot \rho(x_{\pm}, y_{\pm}, t)  + \nabla \cdot (v \rho) = 0,
\end{align}
where the velocity $v$ is interpreted in the Eulerian sense and are given by the right hand side of Eqs.~\eqref{x1}-\eqref{t1} with the sums converted to integrals. They become 
\begin{figure*}[t!]
    \centering
    \includegraphics[width= 1.75 \columnwidth]{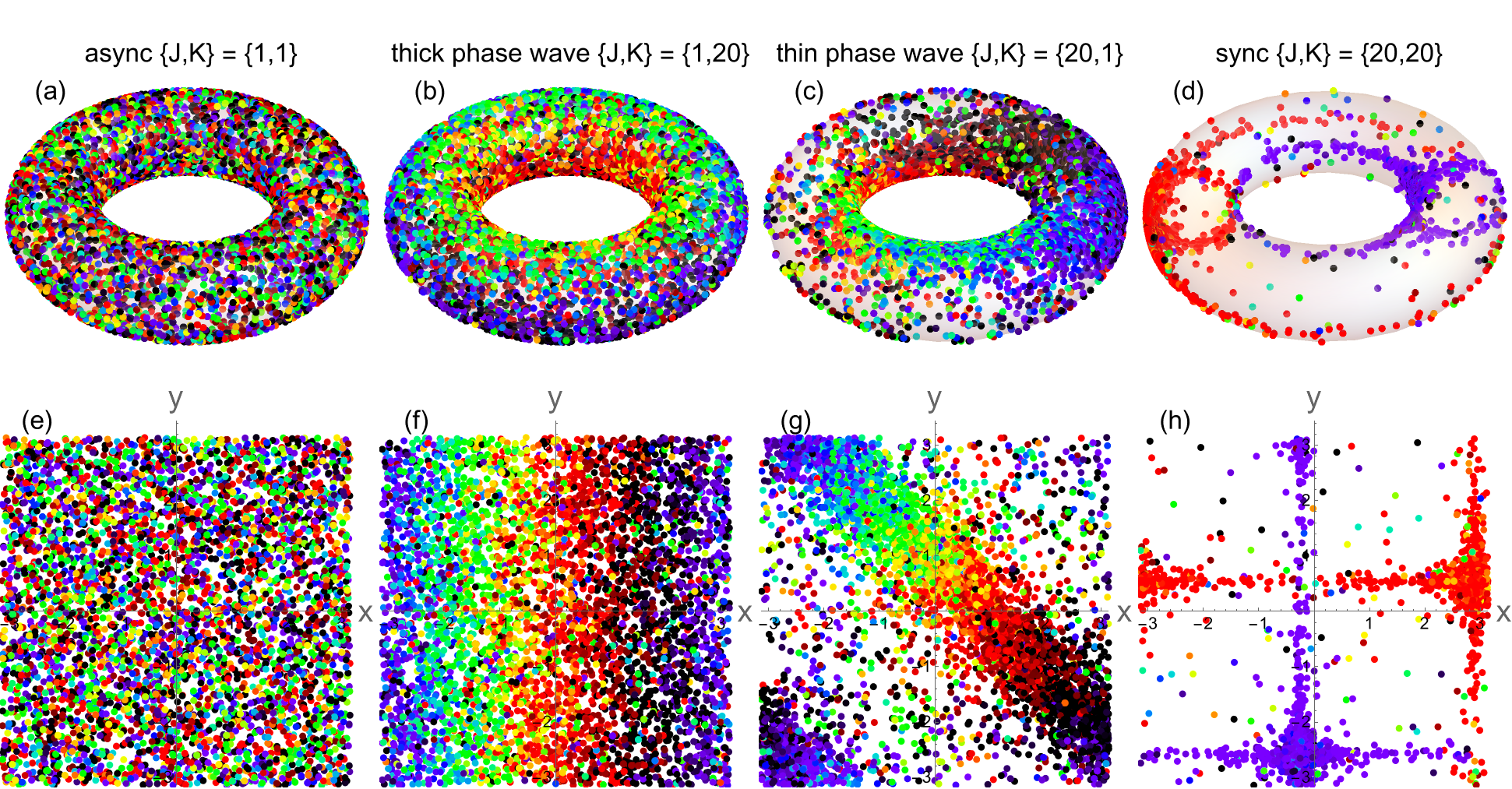}
    \caption{Collective states of the model Eqs.~\eqref{eqx2}-~\eqref{eqtheta2}. Top row, swarmalators are depicted as dots on the torus $(x_i, y_i)$ with the color indicating the phase $\theta_i$. Bottom row, swarmalators depicted in the $(x_i, y_i)$ plane. Simulation details: $(dt,T,N) = (0.1,200,5000)$ using an RK4 solver. (a, e) async for $(J,K)=(1,1)$, (b, f) thick phase wave for $(J,K)=(1,20)$, (c, g) thin phase wave for $(J,K)=(20,1)$, (d, h) sync for $(J,K)=(20,20)$.} 
    \label{states-torus-nonidentical}
\end{figure*}
\begin{align}
    v_{x_+} &= J_+ S_+ \sin ( \phi_+ - x_+) + J_- S_- \sin ( \phi_- - x_-) \nonumber \\
            &\quad + \frac{K}{2} T_+ \sin(\psi_+ - y_+) - \frac{K}{2} T_- \sin(\psi_- - y_-), \\
    v_{x_-} &= J_- S_+ \sin ( \phi_+ - x_+) + J_+ S_- \sin ( \phi_- - x_-) \nonumber \\
            &\quad - \frac{K}{2} T_+ \sin(\psi_+ - y_+) + \frac{K}{2} T_- \sin(\psi_- - y_-), \\
    v_{y_+} &= J_+ T_+ \sin ( \psi_+ - y_+) + J_- T_- \sin ( \psi_- - y_-) \nonumber \\
            &\quad + \frac{K}{2} S_+ \sin(\phi_+ - x_+) - \frac{K}{2} S_- \sin(\phi_- - x_-), \\
    v_{y_-} &= J_- T_+ \sin ( \psi_+ - y_+) + J_+ T_- \sin ( \psi_- - y_-) \nonumber \\
            &\quad - \frac{K}{2} S_+ \sin(\phi_+ - x_+) + \frac{K}{2} S_- \sin(\phi_- - x_-) ,
\end{align}
where the rainbow order parameters \footnote{so called because they take maximal values $1$ in the rainbow-like phase wave states} are
\begin{align}
    W_{\pm} = S_{\pm} e^{i \phi_{\pm}} = \langle e^{ i x_{\pm}} \rangle, \\
    Z_{\pm} = T_{\pm} e^{i \psi_{\pm}} = \langle e^{ i y_{\pm}} \rangle,
\end{align}
and $J_{\pm} = (J \pm K)/2$. Note $\langle \cdot \rangle$ denotes the population average. In $x_{\pm}, y_{\pm}$ coordinates, the model becomes a sum of Kuramoto models which simplifies the analysis. Next we consider perturbations around the async state $\rho(x_{\pm}, y_{\pm},t) = \rho_0 + \epsilon \eta(x_{\pm}, y_{\pm},t)$ and substitute into the continuity equation. The $O(\epsilon)$ term yields
\begin{align}
    \dot{\eta} + \rho_0 (\nabla . v_1) = 0,
\end{align}
where we have plugged $v = v_0 + \epsilon v_1 = 0 + \epsilon v_1 = \epsilon v_1$; that is, $v_0 = 0$ in the async state. After much algebra,
\begin{align}
    \dot{S}_{\pm} \propto J_+ S_{\pm}, \\
    \dot{T}_{\pm} \propto J_+ T_{\pm},
\end{align}
is what we find. From this we see the state loses stability when $J_+ = 0$ which implies
\begin{align}
 K_c = -J, \label{eq.34}
\end{align}
the same $K_c$ as the 1D swarmalator model \cite{o2022collective}.

\textbf{Analysis of thick phase wave}. We can deduce the stability region of the thick phase wave with a squeeze argument. In the absence of bistability between the states, our analysis of thin wave and async state confine the stability region of the thick wave to $ -J \leq K \leq J/2$. Numerics indicate this is true. Of course proving this is an open problem (there could be a parameter region where the thin wave coexists with the thick wave or async state and thus has a different stability threshold).

This completes our analysis of identical swarmalators. Figure~\ref{bif-diagram-model1} summarizes our results in a bifurcation diagram. The thick black lines are theoretical predictions given by Eqs.~\eqref{eq.21} and \eqref{eq.34}.
\section{Nonidentical swarmalators}
Now we consider the harder and more realistic case of swarmalators with distributed natural frequencies
\begin{align}
    \dot{x}_i &= u_i + \frac{J}{N} \sum_j \sin(x_j - x_i)\cos(\theta_j - \theta_i), \label{eqx2} \\
    \dot{y}_i &= v_i + \frac{J}{N} \sum_j \sin(y_j - y_i) \cos(\theta_j - \theta_i), \label{eqy2} \\
    \dot{\theta}_i &= w_i + \frac{K}{N} \sum_j \sin(\theta_j - \theta_i) \left(\cos(x_j - x_i) + \cos(y_j - y_i)\right), \label{eqtheta2} 
\end{align}
where $u,v,w$ are drawn from Lorentzians $g(x)=\Delta/\left[\pi(x^2+\Delta^2)\right]$ with zero center and common width $\Delta$. In sum/difference coordinates these become
\begin{align}
    v_{x_+} &= u_+ + J_+ S_+ \sin ( \phi_+ - x_+) + J_- S_- \sin ( \phi_- - x_-) \nonumber \\
            &\quad + \frac{K}{2} T_+ \sin(\psi_+ - y_+) - \frac{K}{2} T_- \sin(\psi_- - y_-), \\
    v_{x_-} &= u_- + J_- S_+ \sin ( \phi_+ - x_+) + J_+ S_- \sin ( \phi_- - x_-) \nonumber \\
            &\quad - \frac{K}{2} T_+ \sin(\psi_+ - y_+) + \frac{K}{2} T_- \sin(\psi_- - y_-), \\
    v_{y_+} &= v_+ - J_+ T_+ \sin ( \psi_+ - y_+) + J_- T_- \sin ( \psi_- - y_-) \nonumber \\
            &\quad + \frac{K}{2} S_+ \sin(\phi_+ - x_+) - \frac{K}{2} S_- \sin(\phi_- - x_-), \\
    v_{y_-} &= v_- + J_- T_+ \sin ( \psi_+ - y_+) + J_+ T_- \sin ( \psi_- - y_-) \nonumber \\
            &\quad - \frac{K}{2} S_+ \sin(\phi_+ - x_+) + \frac{K}{2} S_- \sin(\phi_- - x_-) ,
\end{align}
where $(u_{\pm}, v_{\pm}) = (u\pm w, v \pm w)$.

Figure~\ref{states-torus-nonidentical} shows this model produces blurry versions of the async, phase waves, and sync states. Like the Kuramoto model, the population splits into locked and drifting subgroups. In the sync state, the slow swarmalators are locked in the clusters, the fast one drifting around it. In the phase waves the split is more complex. Consider the thin phase wave with identical swarmalators $\Delta=0$ defined by $x = y = \theta$ (Fig.~\ref{states-torus}(i)). In sum/difference coordinates this becomes $x_- = y_- =0$. When $\Delta>0$ is turned on, a locked/drifting split occurs in the $x_-, y_-$ directions while the $x_+, y_+$ direction have full rotations. This can be seen in Fig.~\ref{states-torus-nonidentical}(g) where the dense phase wave is surrounded by a sparser `sea' of incoherent drifters. The thick phase wave behaves similarly.
\begin{figure}[t!]
    \centering
    \includegraphics[width= \columnwidth]{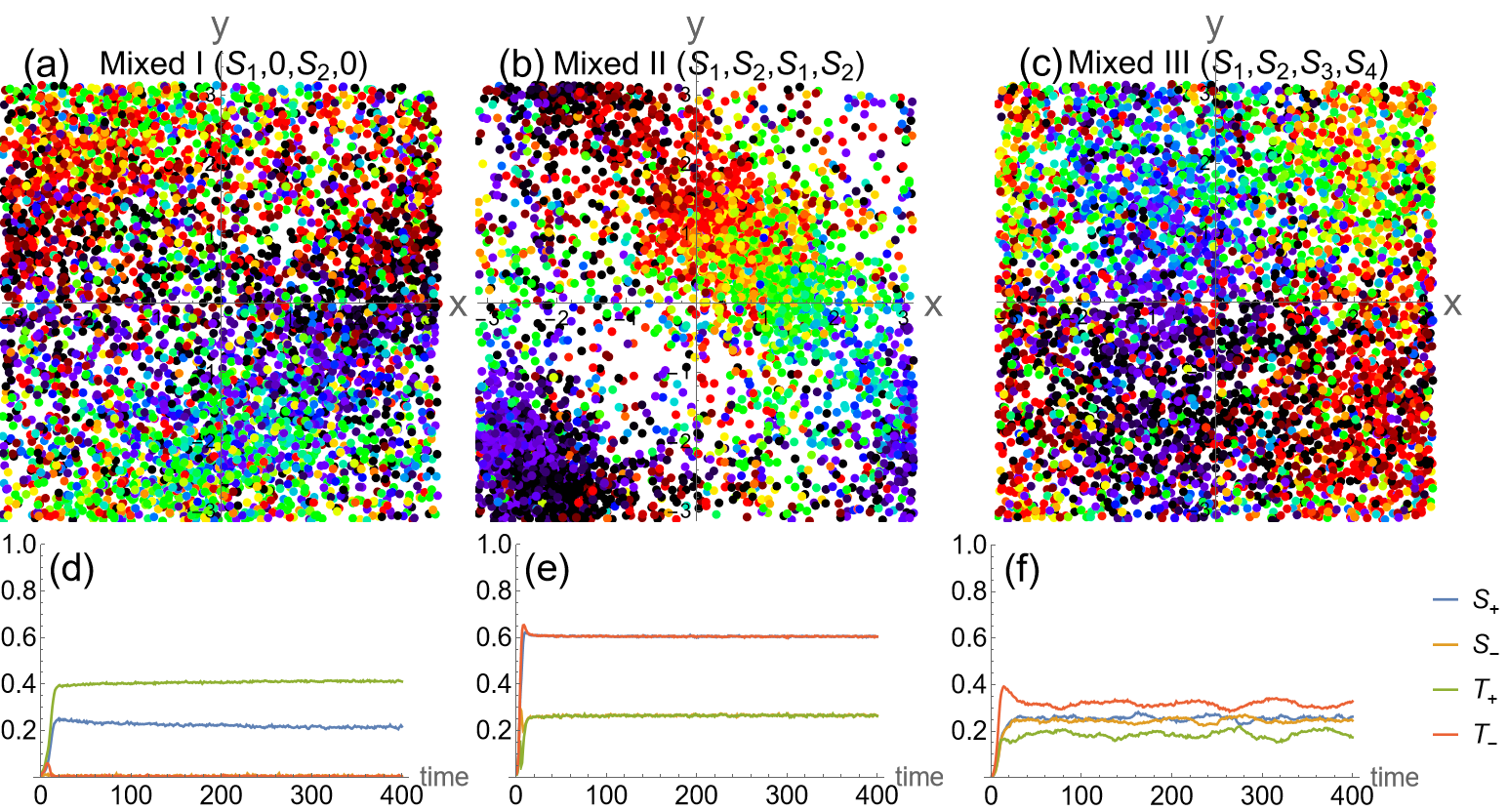}
    \caption{Mixed states. (a) Mixed I for $(J,K)=(15,-4.5)$, (b) mixed II for $(J,K)=(15,1.35)$, and (c) mixed III for $(J,K)=(1.5,10)$. $5000$ swarmalators were randomly chosen from the total $10^5$ swarmalators for these scatter plots. Here, $(dt,T,N)=(0.25,400,10^5)$. Lower panel (d-f) show the variation of order parameters in these states.} 
    \label{mixed}
\end{figure}

Figure~\ref{mixed} shows the model also generates new \textit{mixed} states, which are a blend between the phase waves and the sync states (they fill a link in the bifurcation sequence phase wave $\rightarrow$ mixed $\rightarrow$ sync). It is a challenge to display their structure clearly, but what is happening is the symmetric phase wave develops an asymmetric buckle: two clumps which eventually grow into the sync clusters. Mixed states were found in the 1D swarmalator model and are hard to analyze because the locked/drifting structure, on which the analysis hinges, breaks down. For example in the thin phase wave, swarmalators previously locked in the $x_-, y_-$ directions begin to wobble.

Table~\ref{states} shows the order parameters $(S_+, S_-, T_+, T_-)$ take different values in each state and so can be used to categorize them. We assume without loss of generality that  $S_+, T_+ > S_-,T_-$ and $S_+ > T_+$ which break the degeneracies in the phase waves and mixed states (recall the $(x_+, x_-) \rightarrow (x_-, x_+), (y_+, y_-) \rightarrow (y_-, y_+), (x_+, x_-, y_+, y_-) \rightarrow (y_+, y_-, x_+, x_-)$ symmetries in our model generated clockwise phase waves, counter-clockwise phase waves etc.) Notice that $S_+, S_-$ etc. take different but constant values in the mixed states, which indicates their asymmetric structure. The bottom row of Figure~\ref{mixed} confirms this by showing time series of the order parameters in the mixed states.
\begin{table}
\caption{Order parameter values in each state \label{states}}
\centering
\begin{tabular}{ |m{3cm}||m{1cm}|m{1cm}|m{1cm}|m{1cm}| }
 \hline
 \multicolumn{5}{|c|}{Emerging states} \\
 \hline
 Name& $S_{+}$ &$S_{-}$&$T_{+}$&$T_{-}$\\
 \hline 
 Sync   & $S (> 0)$    & $S (> 0)$ &  $S (> 0)$ & $S (> 0)$\\
 Thin phase wave&   $S (> 0)$  & 0   & $S (> 0)$ & 0 \\
 Thick phase wave & $S (> 0)$ & 0 &  0 & 0 \\
 Async    & 0 & 0 &  0 & 0 \\
 Mixed I&   $S_1$  & $0$ & $S_2$ & $0$ \\
 Mixed II &$S_1$  & $S_2$ & $S_1$ & $S_2$\\
 Mixed III &$S_1$  & $S_2$ & $S_3$ & $S_4$\\
 \hline
\end{tabular}
\end{table}

Now we derive expressions for the order expressions using a generalized OA ansatz \cite{ott2008low}. The Kuramoto model corresponds to synchronization on the unit circle $\theta \in \mathbb{S}^1$ and has an OA ansatz that is a Poisson kernel. Our model corresponds to sync on $(x, y, \theta) \in \mathbb{S}^1 \times \mathbb{S}^1 \times \mathbb{S}^1$, so we search for an OA ansatz that is a product of Poisson kernels. In $(x_{\pm}, y_{\pm})$ coordinates this is
\begin{align}
&\rho(u,v,w,x_{\pm}, y_{\pm},t) = \frac{1}{16 \pi^4} g(u) g(v) g(w) \nonumber \\ & \times \Big[ 1 + \sum_{n=0}^{\infty} \alpha_x^n e^{i n x_+} + c.c \Big] \times [1 + \sum_{m=0}^{\infty} \beta_x^m e^{i m x_-} + c.c. ] \nonumber \\
& \times \Big[ 1 + \sum_{l=0}^{\infty} \alpha_y^l e^{i l y_+} + c.c \Big]  \times [1 + \sum_{p=0}^{\infty} \beta_y^p e^{i p y_-} + c.c. ],
\end{align}
where the Fourier amplitudes are time dependent and depend on the natural frequencies $\alpha_x(t,u,v,w), \beta_x(t,u,v,w), \alpha_y(t,u,v,w), \beta_y(t,u,v,w)$ and c.c. denotes complex conjugates. Plugging this ansatz into the continuity equation and projecting onto $e^{i x_{\pm}}, e^{i y_{\pm}}$ yields
\begin{align}
\dot{\alpha_x} = & -i u_+ \alpha_x +\frac{J_+}{2} \left(W_{+}^* - W_{+} \alpha_x^2\right) \nonumber \\
& + \frac{J_-}{2} \alpha_x \left(W_{-}^* \beta_x^* -W_{-} \beta_x \right) \nonumber \\ 
& +\frac{K}{4} \alpha_x \left(-Z_{-}^* \beta_y^*+Z_{+}^* \alpha_y^*+Z_{-} \beta_y-Z_{+} \alpha_y\right), \label{alphaxdot}\\ 
\dot{\beta_x} = & -i u_- \beta_x+\frac{J_+}{2} \left(W_{-}^*-W_{-} \beta_x^2\right) \nonumber \\
& +\frac{J_-}{2} \beta_x \left(W_{+}^* \alpha_x^*-W_{+} \alpha_x\right) \nonumber \\
& +\frac{K}{4} \beta_x \left(Z_{-}^* \beta_y^*-Z_{+}^* \alpha_y^*- Z_{-} \beta_y+Z_{+} \alpha_y\right), \label{betaxdot} \\ 
\dot{\alpha_y} = & -i v_+ \alpha_y+\frac{J_+}{2} \left(Z_{+}^*-Z_{+} \alpha_y^2\right) \nonumber \\
& +\frac{J_-}{2} \alpha_y \left(Z_{-}^* \beta_y^*-Z_{-} \beta_y\right) \nonumber \\
& +\frac{K}{4} \alpha_y \left(-W_{-}^* \beta_x^*+W_{+}^* \alpha_x^*+W_{-} \beta_x-W_{+} \alpha_x \right), \label{alphaydot} \\ 
\dot{\beta_y} = & -i v_- \beta_y+\frac{J_+}{2} \left(Z_{-}^*-Z_{-} \beta_y^2\right) \nonumber \\
& +\frac{J_-}{2} \beta_y \left(Z_{+}^* \alpha_y^*-Z_{+} \alpha_y \right) \nonumber \\
& +\frac{K}{4} \beta_y \left(W_{-}^* \beta_x^*- W_{+}^* \alpha_x^*- W_{-} \beta_x+ W_{+} \alpha_x\right). \label{betaydot}
\end{align}
Crucially, these only hold on the sub-manifold $|\alpha_x| = |\beta_x| = |\alpha_y| = |\beta_y| = 1$. The higher order mixed harmonics $e^{i( n x_+ + m x_- + l y_+ + p y_-)}$ only close on this sub-manifold. This is the same structure as reported in the 1D swarmalator model \cite{yoon2022sync}. The expressions for the rainbow order parameters become
\begin{align}
    W_+ = S_+ e^{i \phi_+} = \int \alpha_{x}^*(u,v,w) g(u)g(v)g(w) du dv dw ,\label{wplus}\\ 
    W_- = S_- e^{i \phi_-} = \int \beta_{x}^*(u,v,w) g(u)g(v)g(w) du dv dw ,\label{wminus}\\ 
    Z_+ = T_+ e^{i \psi_+} = \int \alpha_{y}^*(u,v,w) g(u)g(v)g(w) du dv dw ,\label{zplus}\\ 
    Z_- = T_- e^{i \psi_-} = \int \beta_{y}^*(u,v,w) g(u)g(v)g(w) du dv dw . \label{zminus}
\end{align}
Now we use the amplitude equation to derive expressions for the order parameters in each state. The strategy is to use Eqs.~\eqref{alphaxdot}-\eqref{betaydot} to find fixed points expressions for $\alpha_{x}$ etc., and then plug those into the integrals for $W_+$ etc. We set $\phi_{\pm} = \psi_{\pm} = 0$ by going to a suitable frame.

\textbf{Thick phase wave}. We seek solutions with $S_+ > 0, S_- = T_+ = T_- = 0$. This implies $\alpha_x = -i (u_+ / (S_+ J_+)) + \sqrt{1 - (u_+ / (S_+ J_+))^2}$. Plugging this into the integral for $S_+$ Eq.~\eqref{wplus} and using the residue theorem we find
\begin{align}
    S_+ &= \sqrt{1-\frac{4 \Delta}{J_+}} = \sqrt{1 - \frac{8 \Delta}{J+K}} .
\end{align}
We see the state bifurcates from zero at
\begin{align}
    J_{+,c} = 4\Delta.
\end{align}

\textbf{Thin phase wave}. Here two of the order parameters $S_+$ and $T_+$ have nonzero equal values and the rest two ($S_-,T_-$) are zero. We look for a solution of Eqs.~\eqref{alphaxdot}-\eqref{zminus} that satisfies $\dot{\alpha}_x=\dot{\alpha}_y=0$, $\dot{\beta}_{x,y} \ne 0$, $W_+ \ne 0$, $Z_+ \ne 0$, and $W_-=Z_-=0$. Solving with these conditions, from Eqs.~\eqref{alphaxdot}-\eqref{zminus}, we get
\begin{align}
    S_+ = T_+ &= \sqrt{1-\frac{16 \Delta  J_+}{\left(3 J_+-J_-\right) \left(J_-+J_+\right)}} \nonumber \\ &= \sqrt{1-\frac{8 \Delta  (J+K)}{J (J+2 K)}} .
\end{align}
This implies a critical threshold
\begin{align}
\frac{16 \Delta  J_+}{\left(3 J_+-J_-\right) \left(J_-+J_+\right)} = 1.
\end{align}

\textbf{Sync}. Finally, we seek solutions with $(S_+, S_-, T_+, T_-) = (S,S,S,S)$. Repeating the same procedure as before yields 
\begin{align}
    S = \sqrt{1 - \frac{\Delta(3J_+-J_-)}{J_+^2-J_-^2}} = \sqrt{1 - \Big( \frac{2\Delta}{J} + \frac{\Delta}{K} \Big) } ,
\end{align}
which implies an existence boundary 
\begin{align}
\frac{\Delta(3J_+-J_-)}{J_+^2-J_-^2} = 1 .
\end{align}

\textbf{Async}. We finish by using the amplitude equation to prove the stability of the async state defined by $(S_+,S_-,T_+,T_-)=(0,0,0,0).$ For $S_{\pm}=0$ and $T_{\pm}=0$, Eqs.~\eqref{alphaxdot}-\eqref{betaydot} have a set of solutions
\begin{align}
    \alpha_{x,0}(u,v,w,t)&=\exp[-i(u+w)t], \label{alphax1}\\
    \beta_{x,0}(u,v,w,t)&=\exp[-i(u-w)t], \label{betax1}\\
    \alpha_{y,0}(u,v,w,t)&=\exp[-i(v+w)t], \label{alphay1}\\
    \beta_{y,0}(u,v,w,t)&=\exp[-i(v-w)t] \label{betay1}.
\end{align}
We consider a small perturbation about the solution given by Eqs.~\eqref{alphax1}-\eqref{betay1},
\begin{align}
   \alpha_x(u,v,w,t)&= \alpha_{x,0}(u,v,w,t)+ \epsilon \alpha_{x,1}(u,v,w,t),\\
   \beta_x(u,v,w,t)&= \beta_{x,0}(u,v,w,t)+ \epsilon \beta_{x,1}(u,v,w,t),\\
   \alpha_y(u,v,w,t)&= \alpha_{y,0}(u,v,w,t)+ \epsilon \alpha_{y,1}(u,v,w,t),\\
   \beta_y(u,v,w,t)&= \beta_{y,0}(u,v,w,t)+ \epsilon \beta_{y,1}(u,v,w,t),
\end{align}
and the perturbed order parameters become $W_+^{(1)}  = \int \alpha_{x,1}^*(u,v,w,t) g(u)g(v)g(w) du dv dw$ etc.
Now, from Eqs.~\eqref{alphaxdot}-\eqref{betaydot} and adopting the arguments used in Ref.~\cite{yoon2022sync} we get,
\begin{align}
    \frac{d W_{\pm}^{(1)}}{dt} &= -(2 \Delta-\frac{1}{2}J_+) W_{\pm}^{(1)},\\
    \frac{d Z_{\pm}^{(1)}}{dt} &= -(2 \Delta-\frac{1}{2}J_+) Z_{\pm}^{(1)}.
\end{align}
From these we get that the async state is stable if $2\Delta - J_+/2 >0$, i.e., $J_+<4\Delta$, consistent with the result from the thick phase wave. 

\textbf{Mixed states}. We were unable to solve for the mixed states, because they do not correspond to any fixed points in $\alpha_x, \alpha_y, \beta_x, \beta_y$. Each $\alpha_x(u,v,w)$ is unsteady, yet somehow the integral over all of these adds up to a stationary $S_{\pm}, T_{\pm}$. The states are somewhat strange, and do not occur in the Kuramoto model, for instance, which retains the locked/drifting structure in each of its collective state. We hope future researchers will be able to shed some light on them.
\begin{figure}[t]
    \centering
    \includegraphics[width= 0.9 \columnwidth]{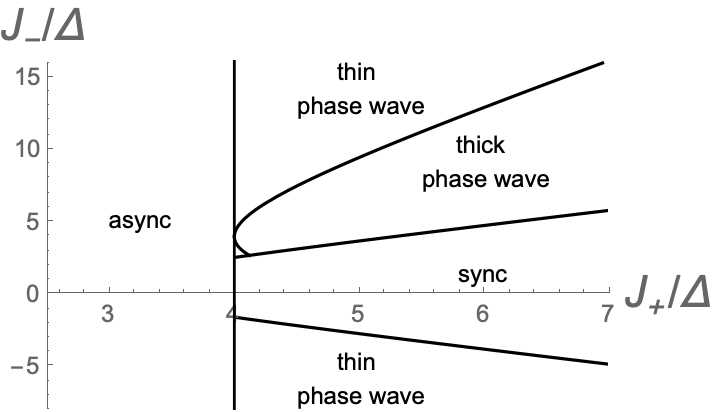}
    \caption{Partial bifurcation diagram for nonidentical swarmalators. Black curves denotes existence boundaries (not stability boundaries) derived theoretically (see text). The exception is the line $J_+ = 4 \Delta$ which is the stability boundary for the async state. Note we have truncated the sync existence boundary at the async stability line. Existence boundaries for the mixed states are unknown, and thus not plotted.} 
    \label{bif-nonid}
\end{figure}
\begin{figure}[t!]
    \centering
    \includegraphics[width= \columnwidth]{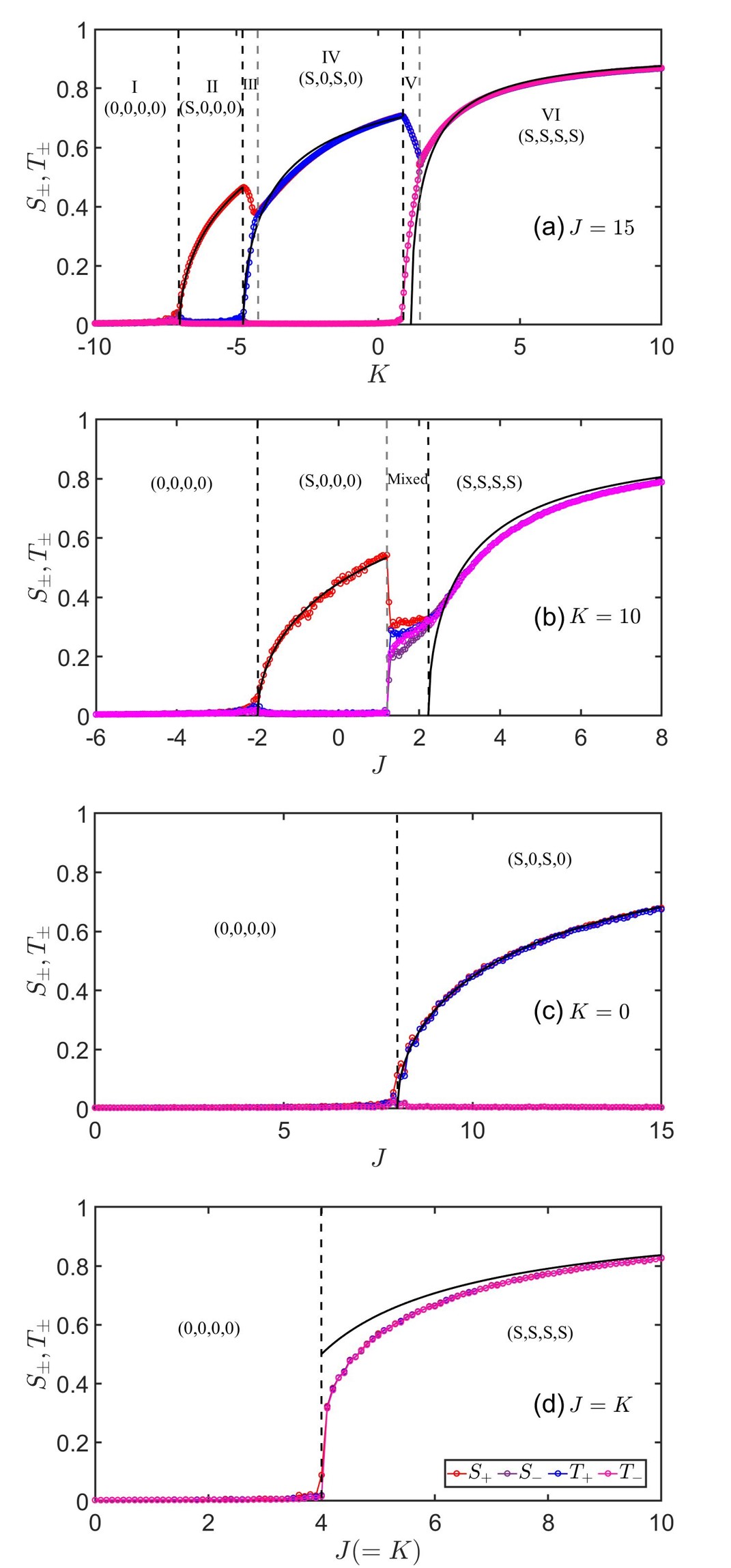}
    \caption{Different bifurcation sequences for nonidentical swarmalators. Black curves are the theoretical predictions. Red, purple, blue, magenta dots for $S_+$, $S_-$, $T_+$, $T_-$, respectively. Sequence of bifurcations occurring for (a) $J=15$ with varying $K$, (b) $K=10$ with varying $J$, (c) $K=0$ with varying $J$, and (d) special case $J=K$. Simulation data: $(T,N) = (250,10^5)$ using an adaptive Julia solver.} 
    \label{bif-seq}
\end{figure}

Figure~\ref{bif-nonid} distills our results with a partial bifurcation diagram in the $(J_+, J_-)$ plane. The line $J_{+}= 4 \Delta$ is a stability boundary, the others are existence boundaries. The existence/stabilities boundaries for the mixed states are not plotted since we could not calculate them. Figure~\ref{bif-seq} reports the different bifurcation sequences the model can sustain by plotting the order parameters versus $K(J)$ for various values of $J(K)$. Panel (a) shows the transition between all six collective states. In panel (b), the thick phase wave is skipped. Panels (c) shows $K=0$, where the system shifts from async to the phase wave only. Panel (d) delineates the special case $J=K$, where the system transitions from async to sync directly. In each panel, our theoretical predictions (thick black curves) match numerics (colored dots) well. An outlier is panel (d), where the match worsens for $K$ near the critical point. We suspect this is a finite $N$ issue.

\section{Discussion and future challenges}
With the overarching goal of developing a set of tools and techniques to analyze swarmalators, we have introduced a simple 2D model with periodic boundary conditions that is tractable. For identical swarmalators, we drew the full bifurcation diagram, for nonidentical swarmalators, a partial bifurcation diagram by deriving expressions for the order parameters $S_{\pm}, T_{\pm}$ in several collective states (the sync and phase waves states) and used those to deduce existence regions.

Missing from our analysis are expressions for $S_{\pm}, T_{\pm}$ in the mixed states, as well as stability proofs for each state (except async, whose stability we found). A big contribution for future work would be to try to fill in these gaps by deriving the evolution equations $\dot{S}_{\pm}, \dot{T}_{\pm}$. We had hoped to do with the OA approach \cite{ott2008low}: apply a time derivative to $S_{\pm}(t) e^{i \psi_{\pm}} = \int \alpha^*(t,u,v,w) g(u) g(v) g(w) du dv dw$, sub in the $\dot{\alpha}$ equation, and use the residue theorem (and same for the $T_{\pm}$; we just use $W_{\pm}$ to keep the example simple). This works for the Kuramoto model order parameter $R e^{i \phi'} = \int \alpha^*(t, \omega) g(\omega) d \omega$. But we couldn't get the contour integrals to close (we had the same problem for the 1D model \cite{yoon2022sync}). So without this tool in hand, how could you derive $\dot{S}_{\pm}, \dot{T}_{\pm}$? A little history might give some hints.

One of the early challenges in the sync field was in fact to derive the analogous evolution equation for the sync order parameter $R$ of the Kuramoto model \cite{strogatz2000kuramoto}. The steady state value of $R$ had famously been found by Kuramoto (for Cauchy distributed natural frequencies this has form $R =\sqrt{\lambda/a}$; the values $\lambda,a$ will arise later), but finding $\dot{R}$ was much harder. Kuramoto and Nishiwaka had some initial theories but they didn't quite work. Later Crawford took up the problem and made progress. He added phase noise to the system $D>0$ which allowed him to do a center manifold calculation around the incoherent state and derive a perturbation series $\dot{R} = -\lambda R - a R^3 + O(R^5)$. Recall this is valid only when $D>0$, but numerics suggested it worked for $D=0$ too. This happened back in 1994. The next big advance came in 2008  when Ott and Antonsen found a way to derive the \textit{full} amplitude equation $\dot{R} = \lambda R -a R^3$ \footnote{which holds under certain conditions}. Strogatz tells the story of all this in \cite{strogatz2000kuramoto} and the more modern parts about OA theory in \cite{lipton2021kuramoto}. Read those to get educated on and engaged by the subject.

This story may guide us to a solution for our problem. Look at all three results at once
\begin{align}
    & R = \sqrt{\frac{\lambda}{a}}, \quad \text{Kuramoto' steady state analysis}. \label{ss} \\
    & \dot{R} = \lambda R -  a R^3 + O(R^5), \hspace{0.1 cm} \text{Crawford's approx } D>0 \label{cr}. \\
    & \dot{R} = \lambda R - a R^3, \quad \text{OA theory, exact for } D \geq 0 . \label{oa}
\end{align}
There are clues here. In order for OA theory Eq.~\eqref{oa} to be consistent with Crawford`s result Eq.~\eqref{cr}, his series must terminate at $O(R^3)$. It's a gripping twist of fate that Crawford didn't calculate the $O(R^5)$ term; if he had, we suspect it must have been zero as $D \rightarrow 0$, which may have enabled him to guess the true $\dot{R} = \lambda R - a R^3$ \textit{without} the OA ansatz that came so much later! In fact, there was another sign that was missed: you might have guessed the $R^5$ term was zero by staring at Kuramoto's steady state expressions for $R$ Eq.~\eqref{ss}. It necessitates an equation of form $\dot{R} = \lambda R -a R^3$ because it must have fixed points $R = \sqrt{\lambda/a}$ and $R=0$ (the $R=0$ one corresponding to the incoherent state).

Crawford or somebody else spotting this would have been a fantastic discovery. Recall how difficult the infinite-$N$ Kuramoto model is: it's a nonlinear, partial, integro-differential equation $\dot{\rho}(\theta, \omega, t) + \partial_{\theta}( \rho(\theta, \omega, t)*(\omega' + K \int \sin(\theta'-\theta) g(\omega') d \omega' d \theta) \rho(\theta, \omega, t)) = 0$. Techniques for analyzing these things are rare -- exact solutions even rarer. And there was Crawford out looking for a modest perturbation series, and then stumbling on the exact solution, like planning a hike to the foot of a mountain, then somehow arriving at the peak itself. It might have meant OA theory, and all the fruits it has borne (solvable models of chimeras \cite{abrams2008solvable}, millenium bridges \cite{strogatz2005crowd}, glass oscillators \cite{ottino2018volcano}, and all the lovely links to the Mobius group \cite{marvel2009identical,engelbrecht2020ott,chen2017hyperbolic}) could have come 14 years early.

Our pitch is to learn from this lesson of history and try to guess $\dot{S}_{\pm}, \dot{T}_{\pm}$ from their steady state values and a perturbation series for them.  For simplicity’s sake consider the 1D swarmalator model from which this paper's model is inspired
\begin{align}
    \dot{x}_i &= \nu_i +  \frac{J}{N} \sum_j \sin(\theta_j -\theta_i) \cos (x_j - x_i), \\
    \dot{\theta}_i &= \omega_i + \frac{K}{N} \sum_j \sin(x_j -x_i) \cos (\theta_j - \theta_i).
\end{align}
This has four collective states: async, phase wave, mixed, and sync. We have expressions for $S_{\pm}$ in 3 of these 4 states, but are missing $S_{\pm}$ in the mixed states and perturbation series for $\dot{S}_{\pm}$
\begin{align}
    &  S_{pw} = \sqrt{1 - \frac{8\Delta}{J+K} }, \label{s1} \\
    &  S_{sync} = \sqrt{1 - 2 \Delta \Big( \frac{1}{J} + \frac{1}{K} \Big) }, \label{s2} \\
    &  S_{mixed,\pm} = ?, \\
    & \dot{S}_+ = ?, \\
    & \dot{S}_- = ?.
\end{align}
If we could fill in the $\dot{S}_{\pm}$ up to $O(S_{\pm}^3)$ by adapting Crawford's center manifold calculation, we might have collected enough jigsaw pieces to infer the full $\dot{S}_{\pm}$. Admittedly, his analysis is already formidable at the level of a single phase $\theta$. Scaling this to a `double phase' $\theta \rightarrow x,\theta$ or triple phase $\theta \rightarrow x,y,\theta$ is daunting (although we have computer algebra systems he didn't have). But if we could pull it off, it would really help this young sub-field on mobile oscillators blossom. We could recreate all the successes the OA ansatz brought to sync studies. 

Why are such questions about swarmalators worth asking in the first place? Why are they important for science? The point is the interplay between sync and self-assembly defines a type of pattern formation that occurs diversely in Nature -- from starfish embryos \cite{tan2022odd} and vinegar eels \cite{quillen2021metachronal} to magnetic domain walls \cite{hrabec2018velocity} and Quincke rollers \cite{liu2021activity} -- yet from a theory perspective is essentially untouched; hardly anyone in the sync community, at least, is asking questions about it. This presents us with an opportunity. It means there might be undiscovered and scientifically useful \cite{o2019review} phenomena hiding under our noses which we are especially equipped to analyze. The value-gain of the 1D and 2D swarmalator models we have proposed is they give us a way to search and solve for these systematically. All you need to do is take your favorite phase model, add in its space-phase partner like we have done (i.e., write down an $\dot{x}$ equation which mirrors the $\dot{\theta}$ equation), run some simulations, and explore what patterns arise. Then these patterns, since they spring from models made of sines and cosines, ought to be tractable with the sync field's technology. To show how many roads you can go down with this approach, we list five examples below.
\begin{figure}[t!]
    \centering
    \includegraphics[width=  0.8\columnwidth]{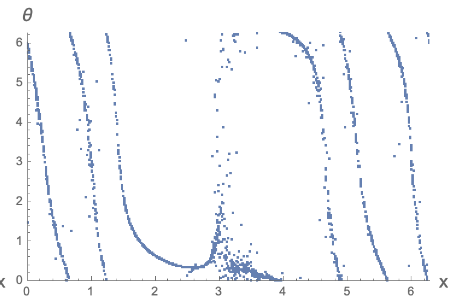}
    \caption{Snapshot of a swarmalator chimera found by simulating model Eqns.~\eqref{chim1},\eqref{chim2} with $(K,J,\alpha) = (-0.025, 1, \pi/2-0.025)$ and $\beta = \alpha$. Simulation parameters $(dt,T,N) = (0.1, 500, 1000)$ using an RK4 solver.} 
    \label{chimera}
\end{figure}

\textit{Chimeras}. The first solvable model of chimeras had form $\dot{\theta}_i = \omega + K \sum_j \sin(\theta_j-\theta_i - \alpha)(1 + \cos(x_j - x_i))$ \cite{abrams2004chimera}. Writing this with its space-phase partner is
\begin{align}
\dot{x}_i = \nu + \frac{J}{N} \sum_j \sin(x_j- x_i - \beta)(1 + \cos(\theta_j - \theta_i)), \label{chim1} \\
\dot{\theta}_i = \omega + \frac{K}{N} \sum_j \sin(\theta_j-\theta_i - \alpha)(1 + \cos(x_j - x_i)). \label{chim2}
\end{align}
Then set $\omega = \nu = 0$ by a change of frame, $J=1$ by rescaling time, and $\beta = \alpha$ for ease. As you sweep the $(K,\alpha)$ plane, might we find a swarmalator chimera? Figure~\ref{chimera} shows a sighting from our early searches. A sync head appears as before, but now the asynchronous part of the chimera is splayed out like a tail. It appears to be non-stationary; the tail thrashes so to speak, splitting from a clean splay wave into a disintegrated mess, and then reforming again (the figure is a snapshot of the state before a thrash). We provide a Mathematica notebook \footnote{https://github.com/Khev/swarmalators/tree/master/1D/on-ring/chimera} where you can simulate it yourself. 

Spiral wave chimeras \cite{shima2004rotating,martens2010solvable}, which are supported by models of the form $\dot{\phi}(\mathbf{x},t) = \omega + \int G(\mathbf{x}' - \mathbf{x}) \sin ( \phi(\mathbf{x'},t) - \phi(\mathbf{x},t) - \alpha) d \mathbf{x}'$ where $\mathbf{x} \in \mathbb{R}^2$, could be generalized in the same way. What might happen here if positions $\mathbf{x}$ are allowed to evolve? Would the incoherent core unstick and slide among the spiral arms?

\textit{Excitable swarmalators}. In 1987, Kuramoto, Shinomoto, and Sakaguchi asked an interesting question \cite{kuramoto1}: can a population of excitatory elements excite themselves into steady oscillations? When uncoupled, each element lies at a fixed point. But they can be excited into an oscillation during which they emit a pulse before falling back to equilibrium. If the elements are now coupled, the pulse can stimulate other elements into oscillations, which may produce more pulses and oscillations in a chain reaction. Kuramoto and his colleagues proved this reaction could indeed be self-sustaining by analyzing a simple model equivalent to $\dot{\theta}_i = \omega_i + b \cos \theta + K/N \sum_j \delta(\theta_j)$ where $\omega_i < b$ guarantees the elements do not oscillate when uncoupled and the $\sum_j \delta(\theta_j)$ term represents pulses fired at $\theta = 0$. Later the model was analyzed fully \cite{o2016dynamics}. 

If the oscillators now swarm around a 1D ring, the pulses ought to get weighed by a distance kernel. A possible setup is
\begin{align}
    \dot{x}_i &= \nu_i + a \cos x_i + \frac{J}{N} \sum_j \cos(x_j - x_i) \delta(\theta_j), \\
    \dot{\theta}_i &= \omega_i + b \cos \theta_i + \frac{K}{N} \sum_j \cos(x_j - x_i) \delta(\theta_j).
\end{align}
We have assumed the the spatial dynamics are excitatory too and have the same pulse coupling, although of course you could play with this structure.

\textit{Winfree model}. The Winfree model \cite{winfree1980geometry} has general form $\dot{\theta}_i = \omega_i + \frac{K}{N} \sum_j R(\theta_i) P(\theta_j)$, where $P(\theta_j)$ models the pulse sent by oscillator $j$, and $R(\theta_i)$ models how the pulse is received. If the oscillators were now running around a 1D ring, perhaps the pulse function might be weighted with a distance function $R(\theta_i) P(\theta_i) G(x_j - x_i)$, with $G(x_j - x_i) = \cos(x_j -x_i)$
\begin{align}
\dot{x}_i &= \nu_i + \frac{J}{N} \sum_j \sin \theta_i (1 + \cos \theta_j) \sin(x_j - x_i),   \\
\dot{\theta}_i &= \omega_i + \frac{K}{N} \sum_j \sin \theta_i (1 + \cos \theta_j) \cos(x_j - x_i).
\end{align}
Here, the spatial dynamics model aggregation, as per the $\sin(x_j-x_i)$ term, altered by the reception of pulses $\sin \theta_i (1 + \cos \theta_j)$ which are \textit{not} weighted by distance kernel to make the model as symmetric as possible (equivalent to regime where the decay length of the pulses in space $l_x$ is far greater than that in phase $l_{\theta}$).

\textit{Vicsek model with non-local coupling}. Vicsek model, one of the model problems from active matter and has form $\dot{x}_i = \hat{n} + D \xi_x(t), \dot{\theta}_i = \omega_i + D \xi_{\theta}(t) \langle \theta_i \rangle_{\mathcal{N}}$ where $\hat{n} = (\cos \theta, \sin \theta)$ is the heading angle, $\xi$ are white noise variables, and $\mathcal{N}$ is the set or neighbours within a given distance. Our idea is to swap the Euclidean distance kernel with our trigonometric one $G(x,y) = 2 + \cos x + \cos y$ which casts the model into a form ripe for the tools in our wheelhouse,
\begin{align}
    \dot{x}_i &= v_0 \cos \theta_i + D \xi_x(t), \\
    \dot{y}_i &= v_0 \sin \theta_i + D \xi_y(t), \\
    \dot{\theta}_i &= \omega_i + D_{\theta} \xi_{\theta}(t) + \frac{K}{N} \sum_{j=1}^{N} \sin(\theta_j - \theta_i) \left[ 2 + \cos(x_j - x_i) \right. \nonumber \\
    &\quad \left. + \cos(y_j - y_i) \right].
\end{align}
Hopefully a steady state might appear and you could leverage Kuramoto's self-consistency analysis to solve for it. That would be a nice way to build a footbridge between the closely related fields of synchronization and active matter.

\textit{Extensions of the 1D swarmalator model}. Finally, there are lots of fresh questions waiting to be asked within the structure of the regular 1D swarmalator (and the regular 2D swarmalator on the torus we have presented in this paper for that matter; though we have not mentioned it explicitly, scaling from the 1D to the 2D model for each of the previously examples should be straightforward). Distributed coupling, phase offsets, and delayed coupling could be input to the model,
\begin{align}
    \dot{x}_i &= \nu_i +  \frac{1}{N} \sum_j J_{ij}\sin(\theta_j(t-\tau) -\theta_i -\alpha) \cos (x_j - x_i), \\
    \dot{\theta}_i &= \omega_i + \frac{1}{N} \sum_j K_{ij}\sin(x_j(t-\tau) -x_i - \beta) \cos (\theta_j - \theta_i).
\end{align}
If $K_{ij}, J_{ij}$ were taken to be Gaussian, we suspect we might see mobile forms of glassiness \cite{daido1992quasientrainment}. Or maybe glassiness would surface in simpler $K_i, K_j$ coupling schemes. Bimodal frequencies distributions $g(\omega), g(\nu)$ would also be fun to explore.  Some explorations of these generalizations have been done for the special case of identical swarmalators \cite{o2022swarmalators,hao2023mixed,hong2023swarmalators,lizarraga2023synchronization,sar2023pinning,sar2023swarmalators,sar2023solvable}. Equivalent problems about non-identical swarmalators are wide open.

\bibliographystyle{apsrev}
\bibliography{ref.bib}


\end{document}